\DeclareFontFamily{OT1}{msb}{}{}
\DeclareFontShape{OT1}{msb}{m}{n}
 {  <5> <6> <7> <8> <9> <10> gen * msbm
      <10.95><12><14.4><17.28><20.74><24.88>msbm10}{}
\DeclareMathAlphabet{\bubble}{OT1}{msb}{m}{n}

\def\bZ{{\bubble Z}}

\documentstyle[12pt]{article}
\parskip=\medskipamount
\begin{document}

\def\l#1#2{\raisebox{.0ex}{$\displaystyle
  \mathop{#1}^{{\scriptstyle #2}\rightarrow}$}}
\def\r#1#2{\raisebox{.0ex}{$\displaystyle
\mathop{#1}^{\leftarrow {\scriptstyle #2}}$}}

\newcommand{\p}[1]{(\ref{#1})}
\newcommand{\sect}[1]{\setcounter{equation}{0}\section{#1}}

\makeatletter
\def\eqnarray{\stepcounter{equation}\let\@currentlabel=\theequation
\global\@eqnswtrue
\global\@eqcnt\z@\tabskip\@centering\let\\=\@eqncr
$$\halign to \displaywidth\bgroup\@eqnsel\hskip\@centering
  $\displaystyle\tabskip\z@{##}$&\global\@eqcnt\@ne
  \hfil$\displaystyle{\hbox{}##\hbox{}}$\hfil
  &\global\@eqcnt\tw@ $\displaystyle\tabskip\z@
  {##}$\hfil\tabskip\@centering&\llap{##}\tabskip\z@\cr}
\@addtoreset{equation}{section}
\makeatother


\renewcommand{\thefootnote}{\fnsymbol{footnote}}
\newpage
\setcounter{page}{0}
\pagestyle{empty}
\begin{flushright}
{October 1998}\\
{ITP-UH-23/98}\\
{JINR E2-98-285}\\
{solv-int/9810009}
\end{flushright}
\vfill

\begin{center}
{\LARGE {\bf Fermionic flows and tau function of the}}\\[0.3cm]
{\LARGE {\bf N=(1$|$1) superconformal Toda lattice hierarchy }}\\[1cm]

{\large O. Lechtenfeld$^{a,1}$ and A. Sorin$^{b,2}$}
{}~\\
\quad \\
{\em {~$~^{(a)}$ Institut f\"ur Theoretische Physik, Universit\"at
Hannover,}}\\
{\em Appelstra\ss{}e 2, D-30167 Hannover, Germany}\\[10pt]
{\em {~$~^{(b)}$ Bogoliubov Laboratory of Theoretical Physics, JINR,}}\\
{\em 141980 Dubna, Moscow Region, Russia}~\quad\\

\end{center}

\vfill

\centerline{{\bf Abstract}}
\noindent
An infinite class of fermionic flows of the $N{=}(1|1)$ superconformal Toda
lattice hierarchy is constructed and their algebraic structure is studied.
We completely solve the semi-infinite $N{=}(1|1)$ Toda lattice and chain
hierarchies and derive their tau functions, which may be relevant for
building supersymmetric matrix models. Their bosonic limit is also discussed.

{}~

{\it PACS}: 02.20.Sv; 02.30.Jr; 11.30.Pb

{\it Keywords}: Completely integrable systems; Toda field theory;
Supersymmetry; Discrete symmetries

\vfill
{\em E-Mail:\\
1) lechtenf@itp.uni-hannover.de\\
2) sorin@thsun1.jinr.dubna.su }
\newpage
\pagestyle{plain}
\renewcommand{\thefootnote}{\arabic{footnote}}
\setcounter{footnote}{0}

\noindent{\bf 1. Introduction.}
Recently the $N{=}(1|1)$ supersymmetric generalization of the Darboux
transformation was proposed, and an infinite class of bosonic solutions
of its symmetry equation was constructed in \cite{ls}. These solutions
generate bosonic flows of the $N{=}(1|1)$ supersymmetric Toda lattice
hierarchy in the same way as their bosonic counterparts -- solutions of
the symmetry equation of the Darboux transformation \cite{dly} --
produce the flows of the bosonic Toda lattice hierarchy.
However, due to the supersymmetry it is obvious that besides bosonic flows
the supersymmetric hierarchy also possesses fermionic ones.
A natural question arises about finding solutions of the
symmetry equation which are responsible for the fermionic flows.

The issue of fermionic flows is also interesting from
a slightly different side. It is a well-known fact by now that particular
reductions of the tau function of the bosonic Toda lattice hierarchy
reproduce partition functions of some  matrix models (for review, see e.g.
\cite{b,m,anp} and references therein). It is reasonable to suspect that
the tau function of the $N{=}(1|1)$ supersymmetric Toda lattice hierarchy may
also be relevant in this respect. In view of the longstanding yet unsolved
problem of constructing supermatrix models which differ non-trivially from
bosonic ones, the knowledge of the super tau function might be an important
advance. In order to derive the complete tau function it is clearly necessary
to know both bosonic and fermionic flows, leading us again to the topic of
fermionic flows.

The present paper addresses both above-stated problems.
In section 2 we construct an infinite class of fermionic flows
of the $N{=}(1|1)$ supersymmetric Toda lattice hierarchy
and derive their algebraic structure. In section 3 we present the general
solution of its reduction -- the $N{=}(1|1)$ semi-infinite Toda lattice
hierarchy --, and derive a superdeterminant representation for its
tau function. We also analyze the tau function of the semi-infinite
$N{=}(1|1)$ Toda chain hierarchy and discuss its bosonic limit.
An appendix illustrates the lowest flow equations
and their solutions in superfield components.

{}~

\noindent {\bf 2. Fermionic flows of the N=(1$|$1) Toda lattice
hierarchy.}
In this section we construct an infinite class of fermionic flows of
the $N{=}(1|1)$ superconformal Toda lattice hierarchy and produce their
algebraic structure.

Our starting point is the $N{=}(1|1)$ supersymmetric generalization
of the Darboux transformation \cite{ls},
\begin{eqnarray}
u_{j+1}\ &=&\ \frac{1}{v_j}\quad,\qquad
v_{j+1}\ =\ v_j\,(D_-D_+\ln v_j-u_jv_j)\quad,
\label{toda1}
\end{eqnarray}
where $u_j\equiv u_j(x^+,\theta^+;x^-,\theta^-)$ and
$v_j\equiv v_j(x^+,\theta^+;x^-,\theta^-)$ are bosonic $N{=}(1|1)$
superfields defined on the lattice, $j \in {\bZ}$,
and $D_{\pm}$ are the $N{=}1$ supersymmetric fermionic covariant derivatives
\begin{eqnarray}
 D_{\pm}\ =\ \frac{\partial}{\partial \theta^{\pm}}+
\theta^{\pm}\frac{{\partial}}{{\partial x^{\pm}}}\quad,
\qquad D^2_{\pm}\ =\ \frac{{\partial}}{{\partial x^{\pm}}}
\ \equiv\ {\partial}_{\pm}\quad,\qquad \{D_+,D_-\}\ =\ 0\quad.
\label{sup}
\end{eqnarray}
Equations \p{toda1} are invariant under global $U(1)$ transformations
which rotate the superfields $u_j$ and $v_j$ in opposite directions.
The composite superfield
\begin{eqnarray}
b_j\ \equiv\ u_j\ v_j
\label{b}
\end{eqnarray}
of length dimension
\begin{eqnarray}
[b_j]\ =\ -1
\label{dimb}
\end{eqnarray}
satisfies the $N{=}(1|1)$ superconformal Toda lattice equation
\begin{eqnarray}
D_-D_+ \ln b_j\ =\ b_{j+1}-b_{j-1}\quad.
\label{toda}
\end{eqnarray}
For this reason, the hierarchy of equations invariant under the
Darboux transformation \p{toda1} we call the $N{=}(1|1)$
superconformal Toda lattice hierarchy.

One of the possible ways of constructing invariant equations is to solve
a corresponding symmetry equation \cite{fl}. In the case under
consideration it reads
\begin{eqnarray}
U_{j+1}\ =\ -\frac{1}{v_{j}^2}V_j \quad, \qquad
V_{j+1}\ =\ \frac{v_{j+1}}{v_j}V_j+v_j\,\Bigl(D_-D_+
(\frac{1}{v_j}V_j)-v_jU_j-u_jV_j\Bigr) \quad,
\label{sym}
\end{eqnarray}
where $V_j$ and $U_j$ are bosonic functionals of the superfields $v_j$ and
$u_j$. Any particular solution $V^{p}_{j}, U^{p}_{j}$ generates an
evolution system of equations involving only the superfields $v_j$ and
$u_j$ defined at the same lattice point
with a bosonic evolution time $t_p$
\begin{eqnarray}
\frac{{\partial}}{\partial t_p}\, v_j\ =\ V^{p}_j \quad,\qquad
\frac{{\partial}}{\partial t_p}\, u_j\ =\ U^{p}_j \quad.
\label{evol}
\end{eqnarray}
By construction\footnote{Let us recall that eq.~\p{sym} is just a result of
differentiating eq.~\p{toda1} with respect to the evolution time $t_p$.},
this is invariant with respect to the discrete transformation \p{toda1}
and, therefore, belongs to the hierarchy as defined above. In other words,
different solutions of the evolution system \p{evol}
(which, actually, are given by pairs of superfields $\{v_j,
u_{j}\}$ with different values for $j$) are related by
the discrete Darboux transformation \p{toda1}. Altogether,
invariant evolution systems form a {\it differential\/} hierarchy, i.e.
a hierarchy of equations involving only superfields at a single lattice
point\footnote{In the case of the one- (two-) dimensional bosonic Toda lattice
the differential hierarchy coincides with the Nonlinear Schr\"odinger
(Davey-Stewartson) hierarchy \cite{bx,lsy,dly}.}. In contrast, the discrete
lattice shift (the Darboux transformation), when added
to the differential hierarchy, generates the {\it discrete\/} $N{=}(1|1)$
superconformal Toda lattice hierarchy. Thus, the
discrete hierarchy appears as a collection of an infinite number of
isomorphic differential hierarchies \cite{bx}.

The symmetry equation \p{sym} represents a complicated nonlinear functional
equation, and its general solution is not known.
For a more complete understanding of the hierarchy structure
and its solutions (tau function) it seems necessary to know
as many solutions of eq.~\p{sym} as possible. Ref.~\cite{ls} addressed
this problem and derived a wide class of bosonic solutions.
However, supersymmetry suggests that eq.~\p{sym} possesses fermionic solutions
as well, and that they are responsible for fermionic flows of the hierarchy.
It turns out that such solutions do in fact exist. We shall demonstrate
that the framework developed in \cite{ls} contains a hidden possibility
for generating fermionic flows.

To explain this observation, we briefly review the approach of ref.~\cite{ls}.

First, the functionals $V_j$ and $U_j$ are consistently represented
in terms of a single bosonic functional $\alpha_{0,j}$,
\begin{eqnarray}
V_j\ =\ -v_j\, \alpha_{0,j} \quad, \qquad U_j\ =\ u_j\,{\alpha}_{0,j-1}
\quad,
\label{evo1}
\end{eqnarray}
in terms of which the symmetry equation \p{sym} becomes
\begin{eqnarray}
D_-D_+\alpha_{0,j}\ =\ b_{j+1}\,({\alpha}_{0,j+1}-\alpha_{0,j})\ +\
b_j\,(\alpha_{0,j}-{\alpha}_{0,j-1}) \quad,
\label{sym1}
\end{eqnarray}
where the superfield $b_j$ is defined by eq.~\p{b}
and constrained by eq.~\p{toda}.

Second, the following recursive chain of substitutions is introduced:
\begin{eqnarray}
\alpha^{\pm}_{p,j}\ =\ \pm D^{-1}_{\mp}(b_{j+p+1} \alpha^{\pm}_{p+1,j}
+ (-1)^{p}\, b_j\, {\alpha}^{\pm}_{p+1,j-1}{})\quad, \qquad p=0,1,2,
...\quad,
\label{rec}
\end{eqnarray}
where $\alpha^{\pm}_{2p,j}$ ($\alpha^{\pm}_{2p+1,j}$) are new bosonic
(fermionic) functionals of length dimensions related as
\begin{eqnarray}
[\alpha_{2p,j}]\ =\ [\alpha^{\pm}_{0,j}]+p \quad, \qquad
[\alpha_{2p-1,j}]\ =\
[\alpha^{\pm}_{0,j}]+p-\frac{1}{2} \quad.
\label{dimal}
\end{eqnarray}
Substituting the functional $\alpha^{\pm}_{0,j}$
in terms of $\alpha^{\pm}_{1,j}$ into eq.~\p{sym1}, the latter reads
\begin{eqnarray}
{\mp}D_{\pm} \alpha^{\pm}_{1,j} + {\alpha}^{\pm}_{1,j} D^{-1}_{\mp}
(b_{j+2}-b_j)\ =\  D^{-1}_{\mp}(b_{j+2}{\alpha}^{\pm}_{1,j}-
b_j\, {\alpha}^{\pm}_{1,j-1}) \quad.
\label{rec1}
\end{eqnarray}
Repeating the same procedure applied to the functional
$\alpha^{\pm}_{1,j}$, i.e. substituting $\alpha^{\pm}_{1,j}$ in terms of
$\alpha^{\pm}_{2,j}$ into eq.~\p{rec1}, the resulting equation for
$\alpha^{\pm}_{2,j}$ takes the following form:
\begin{eqnarray}
&&{\pm}D_{\pm} \alpha^{\pm}_{2,j}+{\alpha}^{\pm}_{2,j} D^{-1}_{\mp}
(b_{j+3}-b_{j+1}+b_{j+2}-b_j)\nonumber\\
&&\ =\ D^{-1}_{\mp}(b_{j+3}
{\alpha}^{\pm}_{2,j+1}-
b_{j+1}\alpha^{\pm}_{2,j}+b_{j+2} \alpha^{\pm}_{2,j}- b_j\,
{\alpha}^{\pm}_{2,j-1}) \quad .
\label{rec2}
\end{eqnarray}
Next, the equation for $\alpha^{\pm}_{3,j}$ emerges,
\begin{eqnarray}
&&{\mp}D_{\pm} \alpha^{\pm}_{3,j}+{\alpha}^{\pm}_{3,j} D^{-1}_{\mp}
(b_{j+4}-b_{j+1}+ b_{j+3}-b_j)\nonumber\\
&&\ =\ D^{-1}_{\mp}(b_{j+4}
{\alpha}^{\pm}_{3,j+1}+
b_{j+1}\alpha^{\pm}_{3,j}-b_{j+3} \alpha^{\pm}_{3,j}-
b_j\, {\alpha}^{\pm}_{3,j-1}) \quad,
\label{rec3}
\end{eqnarray}
and so on.

We now analyze the solutions of the equations
arising in this iterative process. It turns out that, at any given $p$,
those equations possess very simple solutions for $\alpha^{\pm}_{2p,j}$
which, however, translate to very non-trivial solutions for the functional
${\alpha}{}^{\pm}_{0,j}$ via relations \p{rec}.
In turn, ${\alpha}{}^{\pm}_{0,j}$ yields flows via eqs.~\p{evol},
\p{evo1}.

Let us start from the equations for the bosonic functionals
$\alpha^{\pm}_{2p,j}$. They admit the solutions \cite{ls}
\begin{eqnarray}
\alpha^{\pm}_{2p,j}\ =\ 1 \qquad \Rightarrow \qquad
[\alpha^{\pm}_{2p,j}]\ =\ 0 \quad,
\label{recsolb}
\end{eqnarray}
and the recursive procedure may be interrupted at every even step (for the
particular case of $p{=}1$ this solution can be seen from eq.~\p{rec2}).
The corresponding $\alpha^{\pm}_{0,j}$, being expressed in terms
of $\alpha^{\pm}_{2p,j}$ \p{recsolb} via relations \p{rec}, generates the
$p$-th bosonic flow of the hierarchy
\begin{eqnarray}
\frac{\partial}{\partial t^{\pm}_p}\,v_j\ =\ -v_j\,
{\alpha^{\pm}_{0,j}} \quad,\qquad
\frac{\partial}{\partial t^{\pm}_p}\,u_j\ =\ u_j\,
{\alpha}{}^{\pm}_{0,j-1}
\qquad\Rightarrow\qquad [t^{\pm}_p]\ \equiv\ -[\alpha^{\pm}_{0,j}]\ =\
p\quad,
\label{evolb}
\end{eqnarray}
where we have used eqs.~\p{evol}, \p{evo1}, \p{dimal}, and \p{recsolb}.
Although this $\alpha^{\pm}_{0,j}$ depends on all superfields
$v_{j+k}$ and $u_{j+k}$ with $0 \leq k \leq 2p$,
by using eq.~\p{toda1} it can be expressed completely in terms of
the superfields $u_j$ and $v_j$ defined at the same lattice point $j$.
In this way the differential hierarchy of bosonic flows
\p{evolb} is generated (see the discussion after eq.~\p{evol}).
For illustration, we present the first two \cite{ls}:
\begin{eqnarray}
\frac{{\partial}}{{\partial t^{+}_1}}\, v\ =\ v \quad, \qquad
\frac{{\partial}}{{\partial t^{+}_1}}\, u\ =\ u \quad,
\label{bf1}
\end{eqnarray}
\begin{eqnarray}
&&\frac{{\partial}}{{\partial t^{+}_2}}\, v\ =\
+{\partial}^{2}_+ v-2(D_+v)D^{-1}_-{\partial}_+(uv) +
2vD^{-1}_-\Bigl[{\partial}_+(vD_+u)+2uvD^{-1}_-{\partial}_+(uv)\Bigr] \quad,
\nonumber\\[8pt]
&&\frac{{\partial}}{{\partial t^{+}_2}}\, u\ =\
-{\partial}^{2}_+ u -2(D_+u)D^{-1}_-{\partial}_+(uv)
+2uD^{-1}_-\Bigl[{\partial}_+ (uD_+v)- 2uvD^{-1}_-{\partial}_+(uv)\Bigr]
\quad,
\label{bf2}
\end{eqnarray}
where $u\equiv u_j(x^+,\theta^+;x^-,\theta^-)$ and
$v\equiv v_j(x^+,\theta^+;x^-,\theta^-)$.

Concerning eqs.~\p{rec1} and \p{rec3} for the fermionic
functionals $\alpha^{\pm}_{1,j}$ and $\alpha^{\pm}_{3,j}$, respectively,
simple inspection shows that they do not allow for constant Grassmann-odd
solutions. Due to this reason ref.~\cite{ls} concluded that the recursive
procedure cannot be interrupted at an odd step, in distinction to the case
of the bosonic Toda lattice \cite{dly}. As a crucial consequence of this
conclusion there is no place for fermionic flows,
at least not in the framework of this iteration procedure.

However, there is a subtle point in this argument, and
we are going to revise the conclusion.
The argument overlooks the possibility of solutions which are
{\it superfield-independent\/} lattice functions.
Indeed, we find the following solutions of eqs.~\p{sym1} and \p{rec}:
\begin{eqnarray}
\alpha^{\pm}_{2p-1,j}\ =\ {\cal I}_j\,\epsilon \quad, \qquad
[\alpha^{\pm}_{2p-1,j}]\ =\ 0 \quad,
\label{recsolf}
\end{eqnarray}
where $\epsilon$ is a dimensionless fermionic constant and ${\cal I}_j$ is
the simple dimensionless bosonic lattice function
\begin{eqnarray}
{\cal I}_j\ \equiv\ -(-1)^{j}
\label{latf}
\end{eqnarray}
which takes only two values, $+1$ or $-1$, and possesses the
following obvious properties:
\begin{eqnarray}
{\cal I}_{j+1}\ =\
{\cal I}_{j-1}\ =\ - {\cal I}_j \qquad{\rm and}\qquad {\cal I}_{j}^{2}\ =\
1 \quad.
\label{recsolf1}
\end{eqnarray}
Therefore, as in the bosonic case, the recursive procedure
can be interrupted here as well at every odd step.
It remains to show how fermionic flows originate from this background.

This goal in mind, let us represent
the bosonic time derivative entering eq.~\p{evol} in the following form:
\begin{eqnarray}
\frac{{\partial}}{\partial t_p}\, \ =\
\epsilon\, \frac{{\partial}}{\partial {\vartheta}_p} \quad,
\label{evolder}
\end{eqnarray}
defining a fermionic time-derivative
$\frac{{\partial}}{\partial {\vartheta}_p}$. Then, eq.~\p{evol} becomes
\begin{eqnarray}
{\epsilon}\, \frac{{\partial}}{\partial \vartheta^{\pm}_p}\, v_j\
=\ -v_j\, {\alpha^{\pm}_{0,j}},\qquad
{\epsilon}\, \frac{{\partial}}{\partial \vartheta^{\pm}_p}\, u_j\
=\ u_j\, {\alpha}{}_{0,j-1}^{\pm}  \qquad \Rightarrow \qquad
[\vartheta^{\pm}_p]\ \equiv\ - [\alpha^{\pm}_{0,j}]\ =\ p-\frac{1}{2}, ~~~
\label{evolf}
\end{eqnarray}
where $\alpha^{\pm}_{0,j}$ should be expressed in terms
of $\alpha^{\pm}_{2p-1,j}$ \p{recsolf} via relations \p{rec},
and eqs.~\p{evo1}, \p{dimal} and \p{recsolf} have been exploited
to arrive at eqs.~\p{evolf}. The fermionic constant $\epsilon$
enters linearly on both sides of eqs.~\p{evolf},
hence the fermionic flows $\frac{{\partial}}{\partial {\vartheta}_p}$
actually do not depend on ${\epsilon}$. In this context
we remark that $\epsilon$ is an artificial parameter,
which need not be introduced at all. However, without $\epsilon$
it is necessary to consider the quantities
$t_p$, $V^{p}_j$, $U^{p}_j$, $\alpha^{\pm}_{2n,j}$
( $\alpha^{\pm}_{2n+1,j}$ )
entering eqs.~\p{evol}, \p{rec} as fermionic (bosonic)
ones from the beginning. Of course, at the end of the analysis
one arrives at the same result \p{evolf}.

Using eqs.~\p{evolf}, \p{recsolf} and \p{rec} for the
fermionic flows, we elaborate the first two of them,
\begin{eqnarray}
{\cal I}_j\frac{{\partial}}{\partial \vartheta^{+}_1} v
\ =\ -D_+v+2vD^{-1}_-(uv) \quad, \qquad
{\cal I}_j\frac{{\partial}}{\partial \vartheta^{+}_1} u
\ =\ -D_+u-2uD^{-1}_-(uv) \quad,
\label{ff1}
\end{eqnarray}
\begin{eqnarray}
&&{\cal I}_j\frac{{\partial}}{\partial \vartheta^{+}_2} v
= -D_+{\partial}_+v+2({\partial}_+v)D^{-1}_-(uv)+
(D_+v)D^{-1}_-D_+(uv) + vD^{-1}_-\Bigl[u{\partial}_+v + (D_+v)D_+u\Bigr]\ ,
~~~~~ \nonumber\\[8pt]
&&{\cal I}_j\frac{{\partial}}{\partial \vartheta^{+}_2} u
= +D_+{\partial}_+u+2({\partial}_+u)D^{-1}_-(uv)+
(D_+u)D^{-1}_-D_+(uv) + uD^{-1}_-\Bigl[v{\partial}_+u + (D_+u)D_+v\Bigr]\ .
\label{ff2}
\end{eqnarray}

Let us note that the two differential hierarchies arising for the
two different values of ${\cal I}_j$ ($+1$ or $-1$) are actually
isomorphic.
Indeed, one can easily see that they are related by the standard automorphism
which changes the sign of all Grassmann numbers.
Thus, in distinction to the bosonic Toda lattice, where the Darboux
transformation does not change the direction of evolution times in the
differential hierarchy \p{evol}, its supersymmetric counterpart \p{toda1}
reverses the sign of fermionic times in the differential hierarchy. This
supersymmetric peculiarity has no effect on the property
that the supersymmetric {\it discrete\/} hierarchy is a
collection of isomorphic {\it differential\/} hierarchies
like in the bosonic case\footnote{
For the one-dimensional bosonic Toda lattice hierarchy the isomorphism
which relates the differential hierarchies is trivial because
they are identical copies of the single Nonlinear Schr\"odinger hierarchy
\cite{bx}.}.

The flows $\frac{{\partial}}{\partial \vartheta^{-}_k}$ and
$\frac{{\partial}}{\partial t^{-}_k}$ can easily be derived
by applying the invariance transformations
\begin{eqnarray}
{\partial}_{\pm}\ \longrightarrow\ {\partial}_{\mp} \quad, \qquad
D_{\pm}\ \longrightarrow\ \pm D_{\mp}
\label{supaut}
\end{eqnarray}
of the $N{=}(1|1)$ supersymmetry algebra \p{sup} and eqs.~\p{toda1},
\p{toda} and \p{sym1} to the flows $\frac{{\partial}}{\partial
\vartheta^{+}_k}$ \p{ff1}-\p{ff2} and $\frac{{\partial}}{\partial t^{+}_k}$
\p{bf1}-\p{bf2}, respectively, but we do not write them down here.

Using the explicit expressions for the bosonic and fermionic flows
constructed here, one can calculate their algebra
\begin{eqnarray}
&&\Bigl\{\frac{{\partial}}{\partial \vartheta^{\pm}_k}\,,\,
\frac{{\partial}}{\partial \vartheta^{\pm}_l}\Bigr\}\ =\
-2\;\frac{{\partial}}{{\partial t^{\pm}_{k+l-1}}}\quad, \nonumber\\[8pt]
&&\Bigl\{\frac{{\partial}}{\partial \vartheta^{+}_k}\,,\,
\frac{{\partial}}{\partial \vartheta^{-}_l}\Bigr\}\ =\
\Bigl[\frac{{\partial}}{\partial t^{\pm}_k}\,,\,
\frac{{\partial}}{\partial t^{\pm}_l}\Bigr]\ =\
\Bigl[\frac{{\partial}}{\partial t^{+}_k}\,,\,
\frac{{\partial}}{\partial t^{-}_l}\Bigr]\ =\
\Bigl[\frac{{\partial}}{\partial t^{\pm}_k}\,,\,
\frac{{\partial}}{\partial \vartheta^{\pm}_l}\Bigr]\ =\
\Bigl[\frac{{\partial}}{\partial t^{\pm}_k}\,,\,
\frac{{\partial}}{\partial \vartheta^{\mp}_l}\Bigr]\ = 0 \ .~~~~~
\label{alg}
\end{eqnarray}
This algebra coincides with the one used in \cite{i,t},
where the super Toda lattice (STL) hierarchy has been
expressed as a system of infinitely many equations for {\it infinitely many\/}
superfields. Our formulation involves only {\it two independent\/} superfields
($v_j$ and $u_j$). From the point of view of the former approach this
corresponds to extracting those STL hierarchy equations
which can be realized in terms of the superfields $v_j$ and $u_j$ alone
after excluding all other superfields of the STL hierarchy. Keeping
in mind this correspondence it is quite natural to suppose that the
algebra \p{alg} is not only valid for the flows \p{bf1}, \p{bf2},
\p{ff1}--\p{supaut} for which it was in fact calculated,
but for all the other flows as well.
If this is the case, eqs.~\p{alg} may be realized
in the superspace  $\{t^{+}_k, \theta^{+}_k; t^{-}_k,\theta^{-}_k\}$,
\begin{eqnarray}
\frac{{\partial}}{\partial \vartheta^{\pm}_k}\ =\
\frac{\partial}{\partial \theta^{\pm}_k}-
\sum^{\infty}_{l=1}\theta^{\pm}_l
\frac{\partial}{{\partial t^{\pm}_{k+l-1}}} \quad,
\label{covder}
\end{eqnarray}
which is used in what follows.
Here, $\theta^{+}_k$ and $\theta^{-}_k$ are abelian fermionic evolution
times with the dimensions
\begin{eqnarray}
[\theta^{\pm}_k]\ =\ k-\frac{1}{2} \quad.
\label{dim}
\end{eqnarray}

In closing this section we only mention that the flows and
their algebras \p{sup} and \p{alg} admit a consistent reduction to a
one-dimensional subspace by setting
\begin{eqnarray}
{\partial}_{+}\ =\ {\partial}_{-}\ \equiv\ {\partial}
\qquad \Leftrightarrow \qquad
\frac{{\partial}}{{\partial t^{+}_{k}}}\ =\
\frac{{\partial}}{{\partial t^{-}_{k}}}\ \equiv\
\frac{{\partial}}{{\partial t_{k}}} \quad.
\label{red}
\end{eqnarray}
As a result, the $N{=}(1|1)$ supersymmetric Toda chain hierarchy arises,
but its detailed description is beyond the scope of the present work.

{}~

\noindent{\bf 3. The tau function of the semi-infinite N=(1$|$1)
superconformal Toda lattice.}

For the case of the semi-infinite hierarchy, i.e. for the hierarchy
interrupted from the left by the boundary condition
\begin{eqnarray}
u_{-1}\ =\ 0 \quad,
\label{boncond}
\end{eqnarray}
the bosonic and fermionic flows for the
remaining boundary superfield $v_{-1}$ have the extremely
simple form\footnote{To derive these equations it is only necessary to take
into account the $U(1)$ invariance of the flows (consequently, only linear
equations for $v_{-1}$ are admissible at $u_{-1}=0$), the dimensions
\p{evolb} and \p{dim} of bosonic and fermionic times and the algebra \p{alg}.},
\begin{eqnarray}
\frac{{\partial}}{\partial \vartheta^{\pm}_k}\, v_{-1}
\ =\ -D_{\pm}{\partial}^{k-1}_{\pm}v_{-1} \qquad {\rm and} \qquad
\frac{{\partial}}{{\partial t^{\pm}_k}}\, v_{-1}
\ =\ {\partial}^{k}_{\pm} v_{-1} \quad,
\label{eqv}
\end{eqnarray}
and can easily be solved. These equations are consistent
with the algebra \p{alg}, and in its realization \p{covder}
their general solution is
\begin{eqnarray}
v_{-1} = \int\! (\!\prod_{\alpha=\pm}\!\! d\lambda_{\alpha}d\eta_{\alpha})\;
\varphi(\lambda_+, \lambda_-,\eta_+{-}\theta^+,\eta_-{-}\theta^-)\;
\exp{\sum_{\alpha=\pm}\Bigl[(x^{\alpha}{-}
{\eta}_{\alpha}{\theta}^{\alpha}) {\lambda}_{\alpha}+
\sum^{\infty}_{k=1} (t^{\alpha}_k{+}
\eta_{\alpha} {\theta}^{\alpha}_k) {\lambda}^{k}_{\alpha}\Bigr]}~~~~~
\label{gensol}
\end{eqnarray}
where $\varphi$ is an arbitrary function of
bosonic ($\lambda_{\pm}$) and fermionic ($\eta_{\pm}$)
spectral parameters with dimensions
\begin{eqnarray}
[\lambda_{\pm}]\ =\ -1 \quad, \qquad [\eta_{\pm}]\ =\ \frac{1}{2} \quad.
\label{dim1}
\end{eqnarray}

Let us construct the general solution for the superfields
$v_{j}$ and $u_{j}$ at $j \geq 0$.
This can be done by expressing them in terms of the boundary
superfield $v_{-1}$ \p{gensol} via eqs.~\p{toda1} through an
obvious iterative procedure.
We have explicitly checked for the next few values of
$j$ that the resulting expressions, obtained by iteration of
eq.~\p{toda1}, convert to the following nice form:
\begin{eqnarray}
&&v_{2j}\ =\ +(-1)^{j}\frac{\tau_{2j}}{\tau_{2j+1}} \quad\ , \qquad\
v_{2j+1}\ =\ (-1)^{j}\frac{\tau_{2(j+1)}}{\tau_{2j+1}} \quad,
\nonumber\\[8pt]
&&u_{2j}\ =\ -(-1)^{j}\frac{\tau_{2j-1}}{\tau_{2j}} \quad, \qquad
\quad u_{2j+1}\ =\ (-1)^{j}\frac{\tau_{2j+1}}{\tau_{2j}} \quad,
\label{todasol}
\end{eqnarray}
where the $\tau_j$ are\footnote{ The superdeterminant is defined as
${\quad \rm sdet} \left(\begin{array}{cc} A & B \\ C & D
\end{array}\right)\ \equiv\
\det (A-BD^{-1}C ) (\det D)^{-1}$.}
\begin{eqnarray}
&&\tau_0\ \equiv\ -v_{-1} \quad, \qquad
\tau_{2j}\ =\ {\rm sdet}
\biggl(\begin{array}{cc} {\partial}^{p}_+{\partial}^{q}_-\tau_0
& {\partial}^{p}_+{\partial}^{m}_-D_-\tau_0 \\
{\partial}^{k}_+{\partial}^{q}_-D_+\tau_0
& {\partial}^{k}_+{\partial}^{m}_-D_+D_-\tau_0
\end{array}\biggr)^{0\leq p,q \leq j}_{0\leq k,m \leq j-1} \quad,
\nonumber\\[10pt]
&&\tau_{-1}\ =\ 1 \quad, \qquad\
\tau_{2j+1}\ =\ {\rm sdet}
\biggl(\begin{array}{cc} {\partial}^{p}_+{\partial}^{q}_-\tau_0
& {\partial}^{p}_+{\partial}^{m}_-D_-\tau_0 \\
{\partial}^{k}_+{\partial}^{q}_-D_+\tau_0
& {\partial}^{k}_+{\partial}^{m}_-D_+D_-\tau_0
\end{array}\biggr)^{0\leq p,q \leq j}_{0\leq k,m \leq j} \quad.
\label{supdet}
\end{eqnarray}
The supermatrices in eqs.~\p{supdet} can be embedded into a
single supermatrix
\begin{eqnarray}
\Bigl(D^{p}_+D^{q}_-\tau_0\Bigr)_{0\leq p,q\leq M}
\label{supmat}
\end{eqnarray}
with the obvious correspondence. These formulae are plausibly
valid for any value of $j$.

Substituting eqs.~\p{todasol} in to eqs.~\p{toda1} one
obtains the following equation for $\tau_j$:
\begin{eqnarray}
D_-D_+\ln \tau_j\ =\
-\biggl(\frac{\tau_{j-1}}{\tau_{j+1}}\biggr)^{(-1)^j} \quad.
\label{taueq}
\end{eqnarray}
Thus, we see that the general solution \p{todasol} of all equations
belonging to the semi-infinite $N{=}(1|1)$ Toda lattice hierarchy can be
expressed in terms of the single lattice function $\tau_j$ depending via
$\tau_0$ on all hierarchy times. In this respect we can treat
$\tau_j$ as the tau function of the hierarchy. Moreover, this
identification
is supported by the fact that the $\tau_j$ \p{supdet} are in agreement
with a more general expression for the tau function of the STL hierarchy
discussed in \cite{i}.

It is an established fact by now that the tau function of the semi-infinite
bosonic Toda chain hierarchy (restricted by the Virasoro constraints)
reproduces the partition function of the one-matrix model,
which defines two-dimensional minimal conformal matter
interacting with two-dimensional quantum gravity (for review,
see e.g. \cite{m,anp} and references therein). In this context
we are led to consider the reduction \p{red} of the
tau function \p{supdet}, \p{gensol}, and obtain the
tau function of the $N{=}(1|1)$ supersymmetric Toda chain hierarchy.
The latter may be relevant for attacking the old yet unsolved problem of
constructing non-trivial supersymmetric matrix and/or eigenvalue models
\cite{ag,plefka}.
The reduction can be done rather straightforwardly,
and we present the resulting formulae without any comments,
\begin{eqnarray}
&&\tau_{2j}\quad\ =\ {\rm sdet} \biggl(\begin{array}{cc}
{\partial}^{p+q}\tau_0
& {\partial}^{p+m}D_-\tau_0 \\
{\partial}^{k+q}D_+\tau_0
& {\partial}^{k+m}D_+D_-\tau_0
\end{array}\biggr)^{0\leq p,q \leq j}_{0\leq k,m \leq j-1} \quad,
\nonumber\\[10pt]
&&\tau_{2j+1}\ =\ {\rm sdet} \biggl(\begin{array}{cc}
{\partial}^{p+q}\tau_0
& {\partial}^{p+m}D_-\tau_0 \\
{\partial}^{k+q}D_+\tau_0
& {\partial}^{k+m}D_+D_-\tau_0
\end{array}\biggr)^{0\leq p,q \leq j}_{0\leq k,m \leq j} \quad,
\label{supdetred}
\end{eqnarray}
where
\begin{eqnarray}
\tau_{0}\ =\ -\int\!\! d\lambda\; d\eta_+\, d\eta_- \
\varphi(\lambda,\eta_+{-}\theta^+,\eta_-{-}\theta^-)\;
\exp{\Bigl[(x{-}\!\sum_{\alpha=\pm} {\eta}_{\alpha}{\theta}^{\alpha}){\lambda}
+\sum^{\infty}_{k=1} (t_k{+}\!\sum_{\alpha=\pm}
\eta_{\alpha}{\theta}^{\alpha}_k){\lambda}^{k}\Bigr]} \ .~~~~~
\label{gensolred}
\end{eqnarray}
It is also instructive to discuss their bosonic limit which looks as
\begin{eqnarray}
\tau_{2j}|\ =\
\frac{\tau^{T}_{j}[{\rho}_1(\lambda)]}{\tau^{T}_{j-1}[{\rho}_2(\lambda)]}\quad,
\qquad \tau_{2j+1}|\ =\
\frac{\tau^{T}_{j}[{\rho}_1(\lambda)]}{\tau^{T}_{j}[{\rho}_2(\lambda)]}\quad.
\label{supdetred1}
\end{eqnarray}
Here, the vertical line means that all fermionic quantities entering
$\tau_j$ are put equal to zero,
\begin{eqnarray}
{\rho}_1(\lambda)\ \equiv\ \biggl(\frac{{\partial}}{{\partial} \eta_-}
\frac{{\partial}}{{\partial} \eta_+}\;
\varphi \biggr)(\lambda,0,0) \quad,
\qquad {\rho}_2(\lambda)\ \equiv\
{\lambda}^{2}\;\varphi(\lambda,0,0) \quad,
\label{def}
\end{eqnarray}
and $\tau^{T}_{j}[{\rho}(\lambda)]$ denotes the tau function of the
bosonic semi-infinite Toda chain hierarchy with a spectral density
$\rho(\lambda)$. The functions $\tau^{T}_{j}$ can be represented in a
determinant, eigenvalue integral, or matrix integral form \cite{m,anp},
\begin{eqnarray}
\tau^{T}_{j}[{\rho}(\lambda)] &&\ =\
\det\biggl({\partial}^{p+q}\int\!d\lambda
\; {\rho}(\lambda)\; \exp{ \Bigl\{ x\lambda +\sum^{\infty}_{k=1}
t_k{\lambda}^{k}\Bigr\} }\biggr)_{0\leq p,q \leq j} \nonumber\\[8pt]
&&\ \equiv\ \frac{1}{j!}\int \biggl(\prod^{j}_{p=1} d{\lambda}_p \biggr)
\biggl(\prod^{j}_{p\geq q=1}({\lambda}_p{-}{\lambda}_q)^{2}\biggr)\;
\exp{\sum^{j}_{p=1}\Bigl[x{\lambda}_p+\sum^{\infty}_{k=1}
t_k{\lambda}^{k}_p+
\ln{\rho(\lambda_p)}\Bigr]}\nonumber\\[8pt]
&&\ \equiv\ \int\! dM\;
\exp{\;{\rm Tr}\Bigl[xM+\sum^{\infty}_{k=1} t_kM^{k}+\ln{\rho(M)}\Bigr]}\quad,
\label{tautoda}
\end{eqnarray}
where $M$ is an $j\times j$ hermitean matrix. It would be
very interesting to find similar representations (if they exist) for the
supersymmetric tau function $\tau_j$ \p{supdetred}, but we postpone a
discussion of this rather non-trivial problem for future publications.

The form of the bosonic limit \p{supdetred1} of the tau function is not
unexpected because the $N{=}(1|1)$ Toda lattice equation \p{toda}
then reduces to the direct sum of two bosonic Toda lattice
equations with opposite signatures of their kinetic terms.
In eqs.~\p{supdetred1} this
property is in fact reflected by the appearance of two Toda tau functions
raised to opposite powers. Furthermore,
supersymmetry fixes the relative dimensions of their spectral densities,
\begin{eqnarray}
[{\rho}_1(\lambda)]\ =\ [{\rho}_2(\lambda)]+1 \quad,
\label{def1}
\end{eqnarray}
as one can see from eqs.~\p{def} and \p{dim1}. If, in addition, we require
scaling invariance (meaning that only dimensionless constants are allowed)
the spectral densities are forced to obey
\begin{eqnarray}
{\rho}_2(\lambda)\ =\ \lambda\; {\rho}_1(\lambda)
\label{def2}
\end{eqnarray}
modulo an inessential dimensionless factor.
It is a rather curious fact that the two choices
\begin{eqnarray}
\Bigl\{ \quad {\rho}_1(\lambda)=1 \quad,
\quad {\rho}_2(\lambda)=\lambda \quad \Bigr\}
\qquad{\rm or}\qquad
\Bigl\{ \quad {\rho}_1(\lambda)=\frac{1}{\lambda} \quad,
\quad {\rho}_2(\lambda)=1 \quad \Bigr\}
\label{choices}
\end{eqnarray}
yield the partition functions of the one-matrix model ($\rho=1$)
and the one of the generalized Penner model ($\ln\rho=\pm\ln\lambda$) \cite{p}.
In closing we would like to refer also to recent interesting work \cite{anp1},
where a Berezinian construction and similar bosonic limits have been derived
in the context of the reduced Manin-Radul $N{=}1$ supersymmetric  KP hierarchy.

{}~

\noindent{\bf 4. Conclusion.}
In this work we have derived an infinite class of fermionic flows for
the $N{=}(1|1)$ superconformal Toda lattice hierarchy, which are given by
eqs.~\p{evolf}, \p{recsolf} and \p{rec}. Their algebraic structure \p{alg}
has been produced as well. Further, we have constructed the general solution
of the semi-infinite $N{=}(1|1)$ Toda lattice hierarchy and
proposed an explicit expression \p{supdet} for its tau function in a
superdeterminant form. Finally we have obtained the reduced tau function
\p{supdetred} which corresponds to the semi-infinite $N{=}(1|1)$
Toda chain hierarchy. It was seen to have the appropriate bosonic limit
and may be relevant for discovering non-trivial supersymmetric matrix models.

{}~

\noindent{\bf Acknowledgments.}
We would like to thank H. Aratyn, L. Bonora, F. Delduc, V. Kazakov and
A.N. Leznov for their interest in our activity and for useful discussions.
This work was partially supported by the Russian Foundation for Basic
Research, Grant No. 96-02-17634, RFBR-DFG Grant No. 96-02-00180 and by
INTAS grants INTAS-93-127-ext. and INTAS-96-0538.

{}~

\noindent{\bf Appendix.}
For illustration, in this appendix we present the component
form of the fermionic flow equations \p{ff1} and the simplest
nontrivial solution of equations
\p{bf2}, (\ref{ff1}--\ref{ff2}).

In terms of the bosonic ($s,{\overline s},r,{\overline r}$)
and fermionic (${\psi}_{\mp},{\overline {\psi}_{\mp}}$)
components of the superfields $u$ and $v$ defined as
\begin{eqnarray}
&& u \vert=s \quad,\quad
D_{\mp}u \vert={\psi}_{\mp}\quad, \quad
D_-D_+u \vert=r\quad, \nonumber\\
&& v \vert={\overline s}\quad,\quad
D_{\mp}v \vert={\overline {\psi}_{\mp}}\quad, \quad
D_-D_+v \vert={\overline r}\quad,
\label{boscom3}
\end{eqnarray}
where $|$ means the $({\theta}^+,{\theta}^-) \to 0$ limit,
equations \p{ff1} become
\begin{eqnarray}
&&{\cal I}\frac{{\partial}}{\partial \vartheta^{+}_1} {\overline s}
\ =\ -{\overline {\psi}_+}+2{\overline s}
{\partial}^{-1}_-({\psi}_-{\overline s}+s{\overline {\psi}_-})\quad,\quad
{\cal I}\frac{{\partial}}{\partial \vartheta^{+}_1} s
\ =\ -{\psi}_+-2s {\partial}^{-1}_-({\psi}_-{\overline s}+s{\overline
{\psi}_-}) \quad, \nonumber\\
&&{\cal I}\frac{{\partial}}{\partial \vartheta^{+}_1} {\overline r}
\ =\ -{\partial}_+{\overline \psi}_-+2{\overline r}
{\partial}^{-1}_-({\psi}_-{\overline s}+s{\overline {\psi}_-})-
2{\overline \psi}_-{\partial}^{-1}_-(r{\overline s}+s{\overline r}+
{\psi}_-{\overline {\psi}_+}-\!{\psi}_+{\overline {\psi}_-})
-2{\psi}_+{\overline s}^2 -4s{\overline s}{\overline {\psi}_+},
\nonumber\\ &&{\cal I}\frac{{\partial}}{\partial \vartheta^{+}_1} r
\ =\ -{\partial}_+\psi_--2r
{\partial}^{-1}_-({\psi}_-{\overline s}+s{\overline {\psi}_-})+
2 \psi_-{\partial}^{-1}_-(r{\overline s}+s{\overline r}+
{\psi}_-{\overline {\psi}_+}-\!{\psi}_+{\overline {\psi}_-})
+2{\overline {\psi}}_+ s^2 +4s{\overline s}{\psi}_+, \nonumber\\
&&{\cal I}\frac{{\partial}}{\partial \vartheta^{+}_1} {\overline \psi}_-
\ =\ {\overline r}-2{\overline \psi}_-
{\partial}^{-1}_-({\psi}_-{\overline s}+s{\overline {\psi}_-})-
2s{\overline s}^2, \quad
{\cal I}\frac{{\partial}}{\partial \vartheta^{+}_1} \psi_- \ =\ r+2\psi_-
{\partial}^{-1}_-({\psi}_-{\overline s}+s{\overline {\psi}_-})+
2{\overline s}s^2, \nonumber\\
&&{\cal I}\frac{{\partial}}{\partial \vartheta^{+}_1} {\overline \psi}_+
\ =\ {\partial}_+{\overline s}-2{\overline \psi}_+
{\partial}^{-1}_-({\psi}_-{\overline s}+s{\overline {\psi}_-})+
2{\overline s}{\partial}^{-1}_-(r{\overline s}+s{\overline r}+
{\psi}_-{\overline {\psi}_+}-{\psi}_+{\overline {\psi}_-})\quad, \nonumber\\
&&{\cal I}\frac{{\partial}}{\partial \vartheta^{+}_1} \psi_+
\ =\ {\partial}_+s+2\psi_+
{\partial}^{-1}_-({\psi}_-{\overline s}+s{\overline {\psi}_-})-
2s{\partial}^{-1}_-(r{\overline s}+s{\overline r}+
{\psi}_-{\overline {\psi}_+}-{\psi}_+{\overline {\psi}_-})\quad.
\label{compff1}
\end{eqnarray}
Their simplest nontrivial solution as well as a solution of equations
\p{bf2} and \p{ff2} can easily be derived from
general formulae \p{gensol}, (\ref{todasol}--\ref{supdet}) and
it corresponds to $v_{0}, u_{0}$ there. Explicitly, it is:
\begin{eqnarray}
s\ =\ &&- \frac{1}{\tau} \quad, \qquad
{\psi}_{\pm} \ = \ \frac{{\tau_{\pm}}}{{\tau}^2} \quad, \qquad
r \ = \ \frac{{\tau}{\widetilde \tau}-2{\tau_{-}}{\tau_{+}}}{{\tau}^3}\quad,
\nonumber\\[10pt] {\overline s} \ =\ &&-\frac{{\tau}{\widetilde \tau}^2}
{{\tau}{\widetilde \tau}+{\tau_{-}} {\tau_{+}} } \quad\ ,\qquad
{\overline {\psi}_{\pm}} \ = \
-\frac{{\tau}{\widetilde \tau}
({\widetilde \tau}^2{\tau_{\pm}}
\pm {\widetilde \tau}{\tau_{\mp}}{\partial}_{\pm}{\tau}
\mp {\tau}{\widetilde \tau}{\partial}_{\pm}{\tau}_{\mp}
\pm 2{\tau}_+{\tau}_-{\partial}_{\pm}{\tau}_{\mp})}
{({\tau}{\widetilde \tau}+{\tau_{-}} {\tau_{+}})^2} \quad\ , \nonumber\\[10pt]
{\overline r}\ =\ &&\frac{{\widetilde \tau}^3
+2{\tau}{\widetilde \tau}{\partial}_+{\partial}_-{\tau}
+2{\widetilde \tau}
({\tau_{+}}{\partial}_-{\tau_{+}}+{\tau_{-}}{\partial}_+{\tau_{-}})
+2{\tau}({\partial}_-{\tau_{+}}){\partial}_+{\tau_{-}}}
{{\tau}{\widetilde \tau}+{\tau_{-}} {\tau_{+}}}\nonumber\\
&&+\ \frac{{\tau}{\widetilde \tau}^2
({\tau_{+}}{\partial}_-{\tau_{+}}-{\tau_{-}}{\partial}_+
{\tau_{-}}-{\partial}_+{\tau}{\partial}_-{\tau})
+2{\tau}{\widetilde \tau}
({\tau_{-}}({\partial}_-{\tau_{+}}){\partial}_+{\tau}
+{\tau_{+}}({\partial}_+{\tau_{-}}){\partial}_-{\tau})}
{({\tau}{\widetilde \tau}+{\tau_{-}} {\tau_{+}})^2}\nonumber\\
&&+\ 2\frac{{\tau}{\widetilde \tau}^2({\partial}_+{\tau}D_-{\tau})
{\partial}_-{\tau}D_+{\tau}}
{({\tau}{\widetilde \tau}+{\tau_{-}} {\tau_{+}})^3}\quad,
\label{gensolcomp}
\end{eqnarray}
where
\begin{eqnarray}
&&{\tau} \equiv {\tau_{0}} \vert  =   \int\! (\!\prod_{\alpha=\pm}\!\!
d\lambda_{\alpha}d\eta_{\alpha})\; \varphi \;
\exp{\sum_{\alpha=\pm}\Bigl[x^{\alpha} {\lambda}_{\alpha}+
\sum^{\infty}_{k=1} (t^{\alpha}_k {+}\eta_{\alpha} {\theta}^{\alpha}_k)
{\lambda}^{k}_{\alpha}\Bigr]}\;,
\qquad \varphi \equiv -\varphi (\lambda_+,\lambda_-,\eta_+,\eta_-)\;,
\nonumber\\[10pt] &&{\tau_{\pm}} \equiv D_{\pm}{\tau_{0}} \vert  =  \int\!
(\!\prod_{\alpha=\pm}\!\!  d\lambda_{\alpha}d\eta_{\alpha})\;
(\eta_{\pm}{\lambda}_{\pm}
{+}\sum^{\infty}_{k=1} {\theta}^{\pm}_k{\lambda}^{k}_{\pm})\; \varphi \;
\exp\Bigl[{\sum_{\alpha=\pm}(x^{\alpha} {\lambda}_{\alpha}{+}
\sum^{\infty}_{k=1} t^{\alpha}_k {\lambda}^{k}_{\alpha}){+}
\eta_{\mp}\sum^{\infty}_{k=1} {\theta}^{\mp}_k{\lambda}^{k}_{\mp}\Bigr]}\;,
\nonumber\\[10pt]
&& {\widetilde \tau}\equiv D_-D_+{\tau_{0}} \vert  =   \int\!
(\!\prod_{\alpha=\pm}\!\!  d\lambda_{\alpha}d\eta_{\alpha}\;
(\eta_{\alpha}{\lambda}_{\alpha}\!
{+}\sum^{\infty}_{k=1} {\theta}^{\alpha}_k{\lambda}^{k}_{\alpha}))\;
\varphi \; \exp{\sum_{\alpha=\pm}
\Bigl[x^{\alpha} {\lambda}_{\alpha}{+}
\sum^{\infty}_{k=1} t^{\alpha}_k {\lambda}^{k}_{\alpha}}\Bigr]\quad.
\label{taucom}
\end{eqnarray}
At a very particular choice of the function $\varphi$,
this solution corresponds to the one-soliton solution of eqs.
\p{bf2}, (\ref{ff1}--\ref{ff2}), while for general $j$ solutions
(\ref{todasol}--\ref{supdet}) correspond to their $(j{+}1)-$soliton
solutions. Their detailed analysis is out the scope of the present paper.

{}~
\vfill\eject


\begin{thebibliography}{**}

\bibitem{ls}
A.N. Leznov and A.S. Sorin,
{\it Two-dimensional superintegrable mappings and integrable
hierarchies in the $(2|2)$ superspace},
{\sl Phys. Lett.} {\bf B389} (1996) 494, hep-th/9608166;\\
{\it Integrable mappings and hierarchies in the $(2|2)$ superspace},\\
{\sl Nucl. Phys.} (Proc. Suppl.) {\bf B56} (1997) 258.
\bibitem{dly}
V.B. Derjagin, A.N. Leznov and E.A. Yuzbashyan,
{\it Two-dimensional integrable mappings and explicit form of equations
     of $(1{+}2)$-dimensional hierarchies of integrable systems}, \\
IHEP-95-26, MPI 96-39 (1996).
\bibitem{b}
L. Bonora,
{\it Two-matrix models, $W$-algebras and $2D$ gravity},
SISSA-ISAS-170/94/EP.
\bibitem{m}
A. Morozov,
{\it Matrix models as integrable systems}, hep-th/9502091.
\bibitem{anp}
H. Aratyn, E. Nissimov and S. Pacheva,
{\it Constrained KP hierarchies: additional symmetries,
Darboux-B\"acklund solutions and relations to multi-matrix models}, \\
{\sl Int. J. Mod. Phys.} {\bf A12} (1997) 1265, hep-th/9607234.
\bibitem{fl}
A.N. Leznov,
{\it The new look on the theory of integrable systems},
{\sl Physica} {\bf D87} (1995) 48;\\
D.B. Fairlie and A.N. Leznov,
{\it The integrable mapping as the discrete
group of inner symmetry of integrable systems},
{\sl Phys. Lett.} {\bf A199} (1995) 360, hep-th/9305050.
\bibitem{bx}
L. Bonora and C.S. Xiong,
{\it An alternative approach to KP hierarchy in matrix models},
{\sl Phys. Lett.} {\bf B285} (1992) 191, hep-th/9204019;\\
{\it Matrix models without scaling limit},
{\sl Int. J. Mod. Phys.} {\bf A8} (1993) 2973, hep-th/9209041.
\bibitem{lsy}
A.N. Leznov, A.B. Shabat and R.I. Yamilov,
{\it Canonical transformations generated by shifts in nonlinear lattices},
{\sl Phys. Lett.} {\bf A174} (1993) 397.
\bibitem{i}
K. Ikeda,
{\it A supersymmetric extension of the Toda lattice hierarchy},\\
{\sl Lett. Math. Phys.} {\bf 14} (1987) 321.
\bibitem{t}
K. Takasaki,
{\it Differential algebras and D-modules in super Toda lattice hierarchy},\\
{\sl Lett. Math. Phys.} {\bf 19} (1990) 229.
\bibitem{ag}
L. Alvarez-Gaum\'e, H. Itoyama, J.L. Ma\~nes and A. Zadra,
{\it Superloop equations and two dimensional supergravity},
{\sl Int. J. Mod. Phys.} {\bf A7} (1992) 5337, hep-th/9112018.
\bibitem{plefka}
J. Plefka,
{\it Supersymmetric generalizations of matrix models}, PhD thesis,
hep-th/9601041;\\
{\it Iterative solution of the supereigenvalue model},\\
{\sl Nucl. Phys.} {\bf B444} (1995) 333, hep-th/9501120.
\bibitem{p}
R.C. Penner, {\it Perturbation series and the moduli space of
Riemann surfaces},\\
{\sl J. Diff. Geom.} {\bf 27} (1988) 35.
\bibitem{anp1}
H. Aratyn, E. Nissimov and S. Pacheva,\\
{\it Berezinian construction of super-solitons in supersymmetric
     constrained KP hierarchies}, solv-int/9808004;\\
{\it Supersymmetric KP hierarchy: ``ghost'' symmetry structure,
     reductions and Darboux-B\"acklund solutions}, solv-int/9801021.

\end{thebibliography}
\end{document}